\documentstyle[amssymb,preprint,prb,aps]{revtex}

\begin{document}

\begin{titlepage}

\title
{Field-dependent quantum nucleation of antiferromagnetic bubbles}

\author{Rong L\"{u},\footnote {Author to whom 
the correspondence should be addressed.\\
Electronic address: rlu@castu.tsinghua.edu.cn} Yi Zhou,
Jia-Lin Zhu, and Bing-Lin Gu} 
\address{Center for Advanced Study, Tsinghua University,
Beijing 100084, P. R. China}
\date{\today}

\maketitle
\begin{abstract}
The phenomenon of quantum nucleation is studied 
in a nanometer-scale 
antiferromagnet with biaxial symmetry in the presence of a
magnetic field at an arbitrary angle.
Within the instanton approach, we
calculate the dependence of the rate of quantum nucleation
and the crossover temperature on the orientation and strength of the
field for bulk solids and two-dimensional films of antiferromagnets, respectively.
Our results show that the rate of quantum nucleation and the crossover
temperature from thermal-to-quantum transitions depend on the 
orientation and strength of the
field distinctly, which can be tested with the use of existing experimental techniques.

\noindent
{\bf PACS number(s)}:  75.45.+j, 73.40.Gk, 75.30.Gw, 75.50.Ee 
\end{abstract}

\end{titlepage}

One of the most striking manifestations of the quantum character of nature
is quantum tunneling of particles. Following the idea suggested by Caldeira
and Leggett,\cite{1} the tunneling of macroscopic object, known as
Macroscopic Quantum Tunneling (MQT), has become one of the most fascinating
phenomena in condensed matter physics. During the last decade, the problem
of quantum tunneling of magnetization in nanometer-scale magnets has
attracted a great deal of theoretical and experimental interest.\cite{2} The
magnetic MQT includes quantum reversal of the magnetization (or the
N\'{e}el) vector in small single-domain ferromagnets (or antiferromagnets),
quantum nucleation of magnetic bubbles, quantum depinning of domain walls
from defects in bulk magnets, and resonant spin tunneling in molecular
magnetic clusters.\cite{2} MQT in magnetic systems are interesting from a
fundamental point of view as it can extend our understanding of the limits
between quantum and classical physics. On the other hand, MQT is important
to the reliability of small magnetic units in memory devices and the
designing of quantum computers in the future. And the measurement of
magnetic MQT quantities such as the tunneling rates could provide
independent information about microscopic parameters such as the
magnetocrystalline anisotropies and the exchange constants. All this makes
magnetic quantum tunneling an exciting area for theoretical research and a
challenging experimental problem.

The problem of quantum nucleation of a stable phase from a metastable one in
ferromagnetic films is an interesting fundamental problem which allows
direct comparison between theory and experiment.\cite{3} Consider a
ferromagnetic film with its plane perpendicular to the easy axis determined
by the magnetocrystalline anisotropy energy depending on the crystal
symmetry. A magnetic field {\bf H} is applied in a direction opposite to the
initial easy direction of the magnetization {\bf M}, which favors the
reversal of the magnetization. The reversal occurs via the nucleation of a
critical bubble, which then the nucleus does not collapse, but grows
unrestrictedly in volume. If the temperature is sufficiently high, the
nucleation of a bubble is a thermal overbarrier process, and the rate of
thermal nucleation follows the Arrhenius law $\Gamma _T\varpropto \exp
\left( -U/k_BT\right) $, with $k_B$ being the Boltzmann constant and $U$
being the height of energy barrier. In the limit of $T\rightarrow 0$, the
nucleation is purely quantum-mechanical and the rate goes as $\Gamma
_Q\varpropto \exp \left( -{\cal S}_{cl}/\hbar \right) $, with ${\cal S}_{cl}$
being the classical action or the WKB\ exponent which is independent of
temperature. Because of the exponential dependence of the thermal rate on $T$%
, the temperature $T_c$ characterizing the crossover from quantum to thermal
regime can be estimated as $k_BT_c=\hbar U/{\cal S}_{cl}$.

A few theoretical studies of the problem of quantum nucleation have been
around for some time. Privorotskii estimated the exponent in the rate of
quantum nucleation based on the dimensional analysis.\cite{4} Chudnovsky and
Gunther studied the quantum nucleation of a thin ferromagnetic film in a
magnetic field along the opposite direction to the easy axis at zero
temperature by applying the instanton method.\cite{5} Later, Ferrera and
Chudnovsky extended the quantum nucleation to a finite temperature.\cite{6}
Kim studied the effect of an arbitrarily directed magnetic field on the
quantum nucleation of magnetization.\cite{7} The phenomenon of quantum
nucleation was also found in nanometer-scale antiferromagnets,\cite{8,9,10}
where the N\'{e}el vector is the tunneling entity. Theoretical studies on
small single-domain antiferromagnets showed that quantum tunneling should
show up at higher temperatures and higher frequencies than in single-domain
ferromagnets of similar size.\cite{2} This makes nanometer-scale
antiferromagnets more interesting for experimental test.

Up to now theoretical studies on quantum nucleation in antiferromagnets\cite
{8,9,10} were confined to the condition that the metastable state is caused
by the magnetocrystalline anisotropy, which is not easily controlled in
experiments. It is well-known that a magnetic field is a good external
parameter to make the phenomenon of MQT observable. The purpose of this
paper is to extend the previous considerations\cite{8,9,10} to a system with
a magnetic field applied in an arbitrary direction between perpendicular and
opposite to the initial easy axis ($\widehat{z}$ axis). By applying the
instanton method in the spin-coherent-state path-integral representation, we
present the numerical results for the WKB exponent in quantum nucleation of
a thin ferromagnetic film with the magnetic field applied in a range of
angles $\pi /2<\theta _H<\pi $ and $\theta _H=\pi $ respectively, where $%
\theta _H$ is the angle between the initial easy axis ($\widehat{z}$ axis)
and the field. We also discuss the $\theta _H$ dependence of the crossover
temperature $T_c$ from purely quantum nucleation to thermally assisted
processes. Our results show that the distinct angular dependence, together
with the dependence of the WKB\ exponent on the strength of the external
magnetic field, may provide an independent experimental test for quantum
nucleation in an antiferromagnetic film.

For a spin tunneling problem, the rate of magnetization reversal by quantum
tunneling can be determined by the imaginary-time transition amplitude from
an initial state $\left| i\right\rangle $ to a final state $\left|
f\right\rangle $ as 
\begin{equation}
U_{fi}=\left\langle f\right| e^{-{\cal H}T}\left| i\right\rangle =\int {\cal %
D}\left\{ {\bf M}\left( {\bf r},\tau \right) \right\} \exp \left( -{\cal S}%
_E/\hbar \right) ,  \eqnum{1}
\end{equation}
where ${\cal S}_E$ is the Euclidean action which includes the Euclidean
Lagrangian density ${\cal L}_E$ as 
\begin{equation}
{\cal S}_E=\int d\tau d^3{\bf r}{\cal L}_E.  \eqnum{2}
\end{equation}
The system of interest is an antiferromagnet of about 5$\sim $10 nm in
radius at a temperature well below its anisotropy gap. According to the
two-sublattice model,\cite{8} there is a strong exchange energy ${\bf m}%
_1\cdot {\bf m}_2/\chi _{\bot }$ between two sublattices, where ${\bf m}_1$
and ${\bf m}_2$ are the magnetization vectors of the two sublattices with
large, fixed and unequal magnitudes, and $\chi _{\bot }$ is the transverse
susceptibility. In the semiclassical regime and using the
spin-coherent-state path-integral, one gets the Euclidean Lagrangian density
for antiferromagnets (neglecting dissipation with the environment) as\cite
{8,9,10} 
\begin{eqnarray}
{\cal L}_E[\theta ({\bf r},\tau ),\phi ({\bf r},\tau )] &=&i\frac{m_1+m_2}%
\gamma \left( \frac{d\phi }{d\tau }\right) -i\frac m\gamma \left( \frac{%
d\phi }{d\tau }\right) \cos \theta +\frac{\chi _{\bot }}{2\gamma ^2}\left[
\left( \frac{d\theta }{d\tau }\right) ^2\right.  \nonumber \\
&&\left. +\left( \frac{d\phi }{d\tau }\right) ^2\sin ^2\theta \right] +\frac %
12\alpha \left[ (\nabla \theta )^2+(\nabla \phi )^2\sin ^2\theta \right]
+E(\theta ,\phi ),  \eqnum{3}
\end{eqnarray}
where $\gamma $ is the gyromagnetic ratio, $\alpha $ is the exchange
constant,\cite{11} and $\tau =it$ is the imaginary-time variable. The $%
E(\theta ,\phi )$ term includes the magnetocrystalline anisotropy and the
Zeeman energies. $m=m_1-m_2=\hbar \gamma s$, where $s$ is the excess spin
due to the noncompensation of two sublattices. The polar coordinate $\theta $
and the azimuthal coordinate $\phi $ in the spherical coordinate system with 
${\bf l}\cdot \widehat{z}=\cos \theta $, ${\bf l}$ is the N\'{e}el vector of
unit length and $\widehat{z}$ is a unit vector along the ${\bf z}$ axis. The
first term in Eq. (3) is a total imaginary-time derivative, which has no
effect on the classical equations of motion, but it is crucial for the
spin-parity effects.\cite{2,8,12,13,14,15,16,17,18} However, for the closed
instanton trajectory described in this paper (as shown in the following),
this time derivative gives a zero contribution to the path integral, and
therefore can be omitted. In the semiclassical limit, the rate of quantum
nucleation, with an exponential accuracy, is given by 
\begin{equation}
\Gamma _Q\varpropto \exp \left[ -{\cal S}_E^{\min }/\hbar \right] , 
\eqnum{4}
\end{equation}
where ${\cal S}_E^{\min }$ is obtained along the trajectory that minimizes
the Euclidean action ${\cal S}_E$.

In this paper, we study the quantum nucleation of the N\'{e}el vector in
antiferromagnets with biaxial symmetry in the presence of a magnetic field
at arbitrary angles in the $ZX$ plane, which has the following
magnetocrystalline anisotropy energy 
\begin{equation}
E\left( \theta ,\phi \right) =K_1\sin ^2\theta +K_2\sin ^2\theta \sin ^2\phi
-mH_x\sin \theta \cos \phi -mH_z\cos \theta ,  \eqnum{5}
\end{equation}
where $K_1$ and $K_2$ are the longitudinal and the transverse anisotropy
coefficients respectively, and $K_1>0$. In the absence of the magnetic
field, the easy axes of this system are $\pm \widehat{z}$ for $K_1>0$. And
the field is applied in the $ZX$ plane at $\pi /2<\theta _H<\pi $. By
introducing the dimensionless parameters as

\begin{equation}
\overline{K}_2=K_2/2K_1,\overline{H}_x=H_x/H_0,\overline{H}_z=H_z/H_0, 
\eqnum{6}
\end{equation}
Eq. (5) can be rewritten as

\begin{equation}
\overline{E}\left( \theta ,\phi \right) =\frac 12\sin ^2\theta +\overline{K}%
_2\sin ^2\theta \sin ^2\phi -\overline{H}_x\sin \theta \cos \phi -\overline{H%
}_z\cos \theta +\overline{E}_0,  \eqnum{7}
\end{equation}
where $E\left( \theta ,\phi \right) =2K_1\overline{E}\left( \theta ,\phi
\right) $, and $H_0=2K_1/m$. At finite magnetic field, the plane given by $%
\phi =0$ is the easy plane, on which $\overline{E}\left( \theta ,\phi
\right) $ reduces to

\begin{equation}
\overline{E}\left( \theta ,\phi =0\right) =\frac 12\sin ^2\theta -\overline{H%
}\cos \left( \theta -\theta _H\right) .  \eqnum{8}
\end{equation}

We denote $\theta _0$ to be the initial angle and $\theta _c$ the critical
angle at which the energy barrier vanishes when the external magnetic field
is close to the critical value $\overline{H}_c\left( \theta _H\right) $ (to
be calculated in the following). Then, the initial angle $\theta _0$
satisfies $\left[ d\overline{E}\left( \theta ,\phi =0\right) /d\theta
\right] _{\theta =\theta _0}=0$, the critical angle $\theta _c$ and the
dimensionless critical field $\overline{H}_c$ satisfy both $\left[ d%
\overline{E}\left( \theta ,\phi =0\right) /d\theta \right] _{\theta =\theta
_c,\overline{H}=\overline{H}_c}=0$ and $\left[ d^2\overline{E}\left( \theta
,\phi =0\right) /d\theta ^2\right] _{\theta =\theta _c,\overline{H}=%
\overline{H}_c}=0$, which leads to

\begin{eqnarray}
\frac 12\sin \left( 2\theta _0\right) +\overline{H}\sin \left( \theta
_0-\theta _H\right) &=&0,  \eqnum{9a} \\
\frac 12\sin \left( 2\theta _c\right) +\overline{H}_c\sin \left( \theta
_c-\theta _H\right) &=&0,  \eqnum{9b} \\
\cos \left( 2\theta _c\right) +\overline{H}_c\cos \left( \theta _c-\theta
_H\right) &=&0.  \eqnum{9c}
\end{eqnarray}
After some algebra, the dimensionless critical field $\overline{H}_c\left(
\theta _H\right) $ and the critical angle $\theta _c$ are found to be

\begin{eqnarray}
\overline{H}_c &=&\left[ \left( \sin \theta _H\right) ^{2/3}+\left| \cos
\theta _H\right| ^{2/3}\right] ^{-3/2},  \eqnum{10a} \\
\sin \left( 2\theta _c\right) &=&\frac{2\left| \cot \theta _H\right| ^{1/3}}{%
1+\left| \cot \theta _H\right| ^{2/3}}.  \eqnum{10b}
\end{eqnarray}
Then the critical field of this system is $H_c=\overline{H}_c\left(
2K_1/m\right) $, where $m=\hbar \gamma s/V$, and $s$ is the excess spin of
antiferromagnet due to the noncompensation of two sublattices.

It is noted that for the nanometer-scale antiferromagnet at finite magnetic
field, there exists another field $H_{s.f.}$, known as the spin-flop field,
which can rotate the moments of sublattices away from the anisotropy axis.
The spin-flop field is defined as $H_{s.f.}=\sqrt{2H_1H_{ex}}$, with $%
H_1=2K_1/m_1$ being the longitudinal anisotropy field, and $%
H_{ex}=2J_{ex}/m_1$ being the exchange field between two sublattices. $%
m_1=\hbar \gamma S/V$ is the magnetization of one sublattice, where $S$ is
the sublattice spin. Typical values of parameters for the antiferromagnetic
nanoparticle are $K_1\backsim 10^5$ erg/cm$^3$ and $\chi _{\bot }\backsim
10^{-5}$ emu/G$\cdot $cm$^3$. The particle radius is about $12$ nm, the
sublattice spin is $S=2\times 10^5$, and the excess spin is $s=10^3$. It is
easy to obtain that the exchange energy density between sublattices is $%
J_{ex}\left( =\hbar ^2\gamma ^2S^2/V^2\chi _{\bot }\right) \approx 1.9\times
10^{10}$ erg/cm$^3$, $H_0\left( =2K_1/m\right) \approx 9.0\times 10^4$ G,
and $H_{s.f.}\approx 2.8\times 10^5$ G, which shows that the critical field
is smaller than the spin-flop field. Therefore, the small $\epsilon =1-%
\overline{H}/\overline{H}_c$ limit can be performed in calculating the rate
of quantum nucleation of the N\'{e}el vector, at which the two-sublattice
configuration is still valid for antiferromagnets at finite magnetic field.

Now we consider the limiting case that the external applied magnetic field
is slightly lower than the critical field, i.e., $\epsilon =1-\overline{H}/%
\overline{H}_c\ll 1$. At this practically interesting situation, the barrier
height is low and the width is narrow, and therefore the tunneling rate is
large. Introducing $\eta \equiv \theta _c-\theta _0$ $\left( \left| \eta
\right| \ll 1\text{ in the limit of }\epsilon \ll 1\right) $, expanding $%
\left[ d\overline{E}\left( \theta ,\phi =0\right) /d\theta \right] _{\theta
=\theta _0}=0$ about $\theta _c$, and using the relations $\left[ d\overline{%
E}\left( \theta ,\phi =0\right) /d\theta \right] _{\theta =\theta _c,%
\overline{H}=\overline{H}_c}=0$ and $\left[ d^2\overline{E}\left( \theta
,\phi =0\right) /d\theta ^2\right] _{\theta =\theta _c,\overline{H}=%
\overline{H}_c}=0$, Eq. (9a) becomes

\begin{equation}
\sin \left( 2\theta _c\right) \left( \epsilon -\frac 32\eta ^2\right) -\eta
\cos \left( 2\theta _c\right) \left( 2\epsilon -\eta ^2\right) =0. 
\eqnum{11}
\end{equation}
Simple calculations show that $\eta $ is of the order of $\sqrt{\epsilon }$.
Thus the order of magnitude of the second term in Eq. (11) is smaller than
that of the first term by $\sqrt{\epsilon }$ and the value of $\eta $ is
determined by the first term, which leads to $\eta \simeq \sqrt{2\epsilon /3}
$. However, when $\theta _H$ is very close to $\pi /2$ or $\pi $, $\sin
\left( 2\theta _c\right) $ becomes almost zero, and the first term is much
smaller than the second term in Eq. (11). Then $\eta $ is determined by the
second term when $\theta _H\simeq \pi /2$ or $\pi $, which leads to $\eta
\simeq \sqrt{2\epsilon }$ for $\theta _H\simeq \pi /2$ and $\eta \simeq 0$
for $\theta _H\simeq \pi $. Since the first term in Eq. (11) is dominant in
the range of values, $\theta _c$, which satisfies $\tan \left( 2\theta
_c\right) >O\left( \sqrt{\epsilon }\right) $, $\eta \simeq \sqrt{2\epsilon /3%
}$ is valid for $\pi /2+O\left( \sqrt{\epsilon }\right) <\theta _H<\pi
-O\left( \sqrt{\epsilon }\right) $ by using Eq. (10b). Therefore, $\eta
\simeq \sqrt{2\epsilon }$, $0$, and $\sqrt{2\epsilon /3}$ for $\theta
_H\simeq \pi /2$, $\pi $, and $\pi /2+O\left( \sqrt{\epsilon }\right)
<\theta _H<\pi -O\left( \sqrt{\epsilon }\right) $, respectively. In this
case the potential energy $\overline{E}\left( \theta ,\phi \right) $ reduces
to the following equation in the limit of small $\epsilon $,

\begin{equation}
\overline{E}\left( \delta ,\phi \right) =\overline{K}_2\sin ^2\phi \sin
^2\left( \theta _0+\delta \right) +\overline{H}_x\sin \left( \theta
_0+\delta \right) \left( 1-\cos \phi \right) +\overline{E}_1\left( \delta
\right) ,  \eqnum{12}
\end{equation}
where $\delta \equiv \theta -\theta _0$ $\left( \left| \delta \right| \ll 1%
\text{ in the limit of }\epsilon \ll 1\right) $, and $\overline{E}_1\left(
\delta \right) $ is a function of only $\delta $ given by

\begin{equation}
\overline{E}_1\left( \delta \right) =\frac 14\sin \left( 2\theta _c\right)
\left( 3\delta ^2\eta -\delta ^3\right) +\frac 12\cos \left( 2\theta
_c\right) \left[ \delta ^2\left( \epsilon -\frac 32\eta ^2\right) +\delta
^3\eta -\frac 14\delta ^4\right] .  \eqnum{13}
\end{equation}

It can be shown that in the region of $\pi /2+O\left( \sqrt{\epsilon }%
\right) <\theta _H<\pi -O\left( \sqrt{\epsilon }\right) $, $O\left( \sqrt{%
\epsilon }\right) <\theta _c<\pi /2-O\left( \sqrt{\epsilon }\right) $, $\eta 
$ and $\delta $ are of the order of $\sqrt{\epsilon }$, the second term in
Eq. (13) is smaller than the first term in the small $\epsilon $ limit. It
is convenient to use the dimensionless variables 
\begin{equation}
{\bf r}^{\prime }=\epsilon ^{1/4}{\bf r}/r_0,\tau ^{\prime }=\epsilon
^{1/4}\omega _0\tau ,\overline{\delta }=\delta /\sqrt{\epsilon }, 
\eqnum{14a}
\end{equation}
where $r_0=\sqrt{\alpha /2K_1}$, and 
\begin{equation}
\omega _0=\frac{2\gamma K_1}m\left( \frac{\sin \theta _c}{\overline{H}_x+2%
\overline{K}_2\sin \theta _c}+\frac{2\chi _{\bot }K_1}{m^2}\right) ^{-1/2}. 
\eqnum{14b}
\end{equation}
Then the Euclidean action Eq. (2) for $\pi /2+O\left( \sqrt{\epsilon }%
\right) <\theta _H<\pi -O\left( \sqrt{\epsilon }\right) $ becomes 
\begin{eqnarray}
{\cal S}_E\left[ \overline{\delta }\left( {\bf r}^{\prime },\tau ^{\prime
}\right) ,\phi \left( {\bf r}^{\prime },\tau ^{\prime }\right) \right]  &=&%
\frac{r_0^3}{\epsilon \omega _0}\int d\tau ^{\prime }d^3{\bf r}^{\prime
}\left\{ \frac{\chi _{\bot }}{2\gamma ^2}\epsilon ^{3/2}\omega _0^2\left( 
\frac{\partial \overline{\delta }}{\partial \tau ^{\prime }}\right)
^2\right.   \nonumber \\
&&+\frac{\chi _{\bot }}{2\gamma ^2}\epsilon ^{1/2}\omega _0^2\sin ^2\left(
\theta _0+\sqrt{\epsilon }\overline{\delta }\right) \left( \frac{\partial
\phi }{\partial \tau ^{\prime }}\right) ^2-i\frac m\gamma \epsilon \omega
_0\sin \left( \theta _0+\sqrt{\epsilon }\overline{\delta }\right) \phi
\left( \frac{\partial \overline{\delta }}{\partial \tau ^{\prime }}\right)  
\nonumber \\
&&+2K_1\left[ \overline{K}_2^{\prime }\sin ^2\phi \sin ^2\left( \theta _0+%
\sqrt{\epsilon }\overline{\delta }\right) +2\overline{H}_x\sin ^2\left( 
\frac \phi 2\right) \sin \left( \theta _0+\sqrt{\epsilon }\overline{\delta }%
\right) \right.   \nonumber \\
&&+\frac 12\epsilon ^{3/2}\left( \nabla ^{\prime }\overline{\delta }\right)
^2+\frac 12\epsilon ^{1/2}\sin ^2\left( \theta _0+\sqrt{\epsilon }\overline{%
\delta }\right) \left( \nabla ^{\prime }\phi \right) ^2  \nonumber \\
&&\left. \left. +\frac A4\epsilon ^{3/2}\left( \sqrt{6}\overline{\delta }^2-%
\overline{\delta }^3\right) \right] \right\} ,  \eqnum{15}
\end{eqnarray}
where $A=\sin \left( 2\theta _c\right) $. In Eq. (15) we have performed the
integration by part for the term $-i\frac m\gamma \cos \theta \left( \frac{%
d\phi }{d\tau }\right) $ and have neglected the total imaginary-time
derivative. In can be showed that for $\pi /2+O\left( \sqrt{\epsilon }%
\right) <\theta _H<\pi -O\left( \sqrt{\epsilon }\right) $, only small values
of $\phi $ contribute to the path integral, so that one can replace $\sin
^2\phi $ in Eq. (15) by $\phi ^2$ and neglect the term including $\left(
\nabla ^{\prime }\phi \right) ^2$ which is of the order $\epsilon ^2$ while
the other terms are of the order $\epsilon ^{3/2}$. Then the Gaussian
integration over $\phi $ leads to 
\begin{equation}
\int {\cal D}\left\{ \delta \left( {\bf r}^{\prime },\tau ^{\prime }\right)
\right\} \exp \left( -\frac 1\hbar {\cal S}_E^{eff}\right) ,  \eqnum{16}
\end{equation}
where the effective action is 
\begin{eqnarray}
{\cal S}_E^{eff}\left[ \overline{\delta }\left( {\bf r}^{\prime },\tau
^{\prime }\right) \right]  &=&\hbar s\epsilon ^{1/2}r_0^3\left( \frac{\sin
\theta _c}{\overline{H}_x+2\overline{K}_2\sin \theta _c}+\frac{2\chi _{\bot
}K_1}{m^2}\right) ^{1/2}  \nonumber \\
&&\times \int d\tau ^{\prime }d^3{\bf r}^{\prime }\left[ \frac 12\left( 
\frac{\partial \overline{\delta }}{\partial \tau ^{\prime }}\right) ^2+\frac %
12\left( \nabla ^{\prime }\overline{\delta }\right) ^2+\frac A4\left( \sqrt{6%
}\overline{\delta }^2-\overline{\delta }^3\right) \right] ,  \eqnum{17}
\end{eqnarray}
and $s=m/\hbar \gamma $ is the excess spin of antiferromagnets. Introducing
the variables $\overline{\tau }=\tau ^{\prime }\sqrt{A}$ and $\overline{{\bf %
r}}={\bf r}^{\prime }\sqrt{A}$, the effective action Eq. (17) is simplified
as 
\begin{eqnarray}
{\cal S}_E^{eff}\left[ \overline{\delta }\left( \overline{{\bf r}},\overline{%
\tau }\right) \right]  &=&\hbar s\epsilon ^{1/2}r_0^3\frac 1A\left( \frac{%
\sin \theta _c}{\overline{H}_x+2\overline{K}_2\sin \theta _c}+\frac{2\chi
_{\bot }K_1}{m^2}\right) ^{1/2}  \nonumber \\
&&\times \int d\overline{\tau }d^3\overline{{\bf r}}\left[ \frac 12\left( 
\frac{\partial \overline{\delta }}{\partial \overline{\tau }}\right) ^2+%
\frac 12\left( \overline{\nabla }\overline{\delta }\right) ^2+\frac 14\left( 
\sqrt{6}\overline{\delta }^2-\overline{\delta }^3\right) \right] . 
\eqnum{18}
\end{eqnarray}

For the quantum reversal of the N\'{e}el vector in a small particle of
volume $V\ll r_0^3$, the N\'{e}el vector is uniform within the particle and $%
\overline{\delta }$ does not depend on the space $\overline{{\bf r}}$. In
this case Eq. (18) reduces to 
\begin{eqnarray}
{\cal S}_E^{eff}\left[ \overline{\delta }\left( \overline{{\bf r}},\overline{%
\tau }\right) \right]  &=&\hbar s\epsilon ^{5/4}\sqrt{A}V\left( \frac{\sin
\theta _c}{\overline{H}_x+2\overline{K}_2\sin \theta _c}+\frac{2\chi _{\bot
}K_1}{m^2}\right) ^{1/2}  \nonumber \\
&&\times \int d\overline{\tau }\left[ \frac 12\left( \frac{d\overline{\delta 
}}{d\overline{\tau }}\right) ^2+\frac 14\left( \sqrt{6}\overline{\delta }^2-%
\overline{\delta }^3\right) \right] .  \eqnum{19}
\end{eqnarray}
The corresponding classical trajectory satisfies the equation of motion 
\begin{equation}
\frac{d^2\overline{\delta }}{d\overline{\tau }^2}=\frac 12\sqrt{6}\overline{%
\delta }-\frac 34\overline{\delta }^2.  \eqnum{20}
\end{equation}
Eq. (20) has the instanton solution 
\begin{equation}
\overline{\delta }\left( \overline{\tau }\right) =\frac{\sqrt{6}}{\cosh
^2\left( 3^{1/4}\times 2^{-5/4}\overline{\tau }\right) },  \eqnum{21}
\end{equation}
corresponding to the variation of $\delta $ from $\delta =0$ at $\tau
=-\infty $, to $\delta =\sqrt{6\epsilon }$ at $\tau =0$, and then back to $%
\delta =0$ at $\tau =\infty $. The associated classical action is found to
be 
\begin{equation}
{\cal S}_{cl}=\frac{2^{17/4}\times 3^{1/4}}5\hbar s\epsilon ^{5/4}\frac{%
\left| \cot \theta _H\right| ^{1/6}}{\sqrt{1+\left| \cot \theta _H\right|
^{2/3}}}\left( \frac{1+\left| \cot \theta _H\right| ^{2/3}}{1-\epsilon +2%
\overline{K}_2\left( 1+\left| \cot \theta _H\right| ^{2/3}\right) }+\frac{%
2\chi _{\bot }K_1}{m^2}\right) ^{1/2}.  \eqnum{22}
\end{equation}
In the WKB approximation, the tunneling rate $\Gamma $ of a particle
escaping from a metastable state has the relation $\Gamma \varpropto \exp
\left( -B/\hbar \right) $. The WKB exponent $B$ is approximately given by $%
U/\omega _b$, where $U$ is the height of barrier, and $\omega _b$ is the
frequency of small oscillations around the minimum of the inverted potential
and characterizes the width of the barrier hindering the decay process. For
magnetic quantum tunneling, $\omega _b^2\left( \equiv -\overline{E}%
_1^{\prime \prime }\left( \delta _m\right) /M\right) $ is inversely
proportional to the effective mass of the magnets, where the mass is induced
by the transverse component of magnetic field, and $\delta _m$ corresponds
to the position of the minimum of the inverted potential. For general case,
the WKB exponent $B$ should be proportional to the power of the parameter $%
\epsilon =1-H/H_c$ since the height and the width of barrier are
proportional to the power of $\epsilon $. Simple analysis of Eq. (8) shows
that the value of $\epsilon $ should be small in order to obtain a large
tunneling rate. For this case, we obtain the height of barrier as $\overline{%
E}_1\left( =U/2K_1V\right) =\sin \left( 2\theta _c\right) \left( 6\epsilon
\right) ^{3/2}/27$ at $\delta _m=2\sqrt{6\epsilon }/3$ and the oscillation
frequency around the minimum of the inverted potential $-\overline{E}%
_1\left( \delta \right) $ as $\omega _b=2^{-1/4}\times 3^{1/4}\epsilon
^{1/4}\omega _0\sqrt{\sin \left( 2\theta _c\right) }$, where $\omega _0$ is
shown in Eq. (14b). Then we approximately obtain the WKB exponent $B$ as 
\begin{eqnarray}
B &\sim &\frac U{\omega _b}  \nonumber \\
&=&\frac{2^{9/4}}{3^{7/4}}\hbar s\epsilon ^{5/4}\frac{\left| \cot \theta
_H\right| ^{1/6}}{\sqrt{1+\left| \cot \theta _H\right| ^{2/3}}}\left( \frac{%
1+\left| \cot \theta _H\right| ^{2/3}}{1-\epsilon +2\overline{K}_2\left(
1+\left| \cot \theta _H\right| ^{2/3}\right) }+\frac{2\chi _{\bot }K_1}{m^2}%
\right) ^{1/2},  \eqnum{23}
\end{eqnarray}
which agrees up to the numerical factor with the result in Eq. (22) obtained
by using the explicit instanton solution.

For a large non-compensation $\left( m\gg \sqrt{\chi _{\bot }K_1}\right) $,
Eq. (23) reduces to 
\begin{equation}
{\cal S}_{cl}^{FM}=\frac{2^{17/4}\times 3^{1/4}}5\hbar s\epsilon ^{5/4}V%
\frac{\left| \cot \theta _H\right| ^{1/6}}{\sqrt{1-\epsilon +2\overline{K}%
_2\left( 1+\left| \cot \theta _H\right| ^{2/3}\right) }},  \eqnum{24}
\end{equation}
which agrees well with the result of quantum tunneling of magnetization in
single-domain ferromagnetic particles with biaxial symmetry in an
arbitrarily directed field.\cite{19} For a small non-compensation $\left(
m\ll \sqrt{\chi _{\bot }K_1}\right) $, Eq. (23) reduces to the result for
single-domain antiferromagnetic particles, 
\begin{equation}
{\cal S}_{cl}^{AFM}=\frac{2^{17/4}\times 3^{1/4}}5\frac{\sqrt{\chi _{\bot
}K_1}}\gamma V\epsilon ^{5/4}\frac{\left| \cot \theta _H\right| ^{1/6}}{%
\sqrt{1+\left| \cot \theta _H\right| ^{2/3}}},  \eqnum{25}
\end{equation}
which is in good agreement with the result in Refs. 20 and 21.

Now we turn to the nonuniform problem. In case of a thin film of thickness $%
h $ less than the size $r_0/\epsilon ^{1/4}$ of the critical nucleus and its
plane is perpendicular to the initial easy axis, we obtain the effective
action after performing the integration over the $\overline{z}$ variable in
Eq. (19), 
\begin{eqnarray}
{\cal S}_E^{eff}\left[ \overline{\delta }\left( \overline{{\bf r}},\overline{%
\tau }\right) \right] &=&\hbar s\epsilon ^{3/4}r_0^2h\sqrt{\frac 1A}\left( 
\frac{\sin \theta _c}{\overline{H}_x+2\overline{K}_2\sin \theta _c}+\frac{%
2\chi _{\bot }K_1}{m^2}\right) ^{1/2}  \nonumber \\
&&\times \int d\overline{\tau }d^2\overline{{\bf r}}\left[ \frac 12\left( 
\frac{\partial \overline{\delta }}{\partial \overline{\tau }}\right) ^2+%
\frac 12\left( \overline{\nabla }\overline{\delta }\right) ^2+\frac 14\left( 
\sqrt{6}\overline{\delta }^2-\overline{\delta }^3\right) \right] . 
\eqnum{26}
\end{eqnarray}
At zero temperature the classical solution of the effective action Eq. (26)
has $O\left( 3\right) $ symmetry in two spatial plus one imaginary time
dimensions. Therefore, the solution $\overline{\delta }$ is a function of $u$%
, where $u=\left( \overline{\rho }^2+\overline{\tau }^2\right) ^{1/2}$, and $%
\overline{\rho }=\left( \overline{x}^2+\overline{y}^2\right) ^{1/2}$ is the
normalized distance from the ${\bf z}$ axis. Now the effective action Eq.
(26) becomes 
\begin{eqnarray}
{\cal S}_E^{eff}\left[ \overline{\delta }\left( \overline{{\bf r}},\overline{%
\tau }\right) \right] &=&4\pi \hbar s\epsilon ^{3/4}r_0^2h\sqrt{\frac 1A}%
\left( \frac{\sin \theta _c}{\overline{H}_x+2\overline{K}_2\sin \theta _c}+%
\frac{2\chi _{\bot }K_1}{m^2}\right) ^{1/2}  \nonumber \\
&&\times \int duu^2\left[ \frac 12\left( \frac{d\overline{\delta }}{du}%
\right) ^2+\frac 14\left( \sqrt{6}\overline{\delta }^2-\overline{\delta }%
^3\right) \right] .  \eqnum{27}
\end{eqnarray}
The corresponding classical trajectory satisfies the equation of motion 
\begin{equation}
\frac{d^2\overline{\delta }}{du^2}+\frac 2u\frac{d\overline{\delta }}{du}=%
\frac{\sqrt{6}}2\overline{\delta }-\frac 34\overline{\delta }^2.  \eqnum{28}
\end{equation}
By applying the similar method,\cite{5,7} the instanton solution of Eq. (28)
can be found numerically and is illustrated in Fig. 1. The maximal rotation
of the N\'{e}el vector is $\overline{\delta }_{\max }\approx 6.8499$ at $%
\overline{\tau }=0$ and $\overline{\rho }=0$. Numerical integration in Eq.
(27), using this solution, gives the rate of quantum nucleation for a thin
antiferromagnetic film as 
\begin{eqnarray}
\Gamma _Q &\varpropto &\exp \left( -{\cal S}_E/\hbar \right)  \nonumber \\
&=&\exp \left\{ -74.39s\epsilon ^{3/4}r_0^2h\frac{\sqrt{1+\left| \cot \theta
_H\right| ^{2/3}}}{\left| \cot \theta _H\right| ^{1/6}}\right.  \nonumber \\
&&\left. \times \left( \frac{1+\left| \cot \theta _H\right| ^{2/3}}{%
1-\epsilon +2\overline{K}_2\left( 1+\left| \cot \theta _H\right|
^{2/3}\right) }+\frac{2\chi _{\bot }K_1}{m^2}\right) ^{1/2}\right\} . 
\eqnum{29}
\end{eqnarray}
For a large non-compensation, Eq. (29) reduces to the result for quantum
nucleation in a thin ferromagnetic film 
\begin{eqnarray}
\Gamma _Q &\varpropto &\exp \left( -{\cal S}_E^{FM}/\hbar \right)  \nonumber
\\
&=&\exp \left\{ -74.39s\epsilon ^{3/4}r_0^2h\frac{1+\left| \cot \theta
_H\right| ^{2/3}}{\left| \cot \theta _H\right| ^{1/6}\sqrt{1-\epsilon +2%
\overline{K}_2\left( 1+\left| \cot \theta _H\right| ^{2/3}\right) }}\right\}
.  \eqnum{30}
\end{eqnarray}
For a small non-compensation, Eq. (29) reduces to the result for quantum
nucleation in a thin antiferromagnetic film 
\begin{eqnarray}
\Gamma _Q &\varpropto &\exp \left( -{\cal S}_E^{AFM}/\hbar \right)  \nonumber
\\
&=&\exp \left\{ -105.2\frac{\sqrt{\chi _{\bot }K_1}}\gamma \epsilon
^{3/4}r_0^2h\frac{\sqrt{1+\left| \cot \theta _H\right| ^{2/3}}}{\left| \cot
\theta _H\right| ^{1/6}}\right\} .  \eqnum{31}
\end{eqnarray}

At high temperature, the nucleation of the N\'{e}el vector is due to thermal
activation, and the rate of nucleation follows $\Gamma _T\varpropto \exp
\left( -W_{\min }/k_BT\right) $, where $W_{\min }$ is the minimal work
necessary to produce a nucleus capable of growing. In this case the
instanton solution becomes independent of the imaginary-time variable $%
\overline{\tau }$. In order to obtain $W_{\min }$, we consider the effective
potential of the system 
\begin{equation}
U_{eff}=\int d^3{\bf r}\left[ \frac \alpha 2\left( \left( \nabla \theta
\right) ^2+\sin ^2\theta \left( \nabla \phi \right) ^2\right) +E\left(
\theta ,\phi \right) \right] .  \eqnum{32}
\end{equation}
For a cylindrical bubble Eq. (32) becomes 
\begin{equation}
U_{eff}=4\pi K_1\epsilon r_0^2\int_0^\infty d\overline{\rho }\overline{\rho }%
\left[ \frac 12\left( \frac{d\overline{\delta }}{d\overline{\rho }}\right)
^2+\frac 14\left( \sqrt{6}\overline{\delta }^2-\overline{\delta }^3\right)
\right] .  \eqnum{33}
\end{equation}
From the saddle point of the functional the shape of the critical nucleus
satisfies 
\begin{equation}
\frac{d^2\overline{\delta }}{d\overline{\rho }^2}+\frac 1{\overline{\rho }}%
\frac{d\overline{\delta }}{d\overline{\rho }}=\frac{\sqrt{6}}2\overline{%
\delta }-\frac 34\overline{\delta }^2.  \eqnum{34}
\end{equation}
The solution can be found by numerical method similar to the one in Refs. 5
and 7. Fig. 2 shows the shape of the critical bubble in thermal nucleation,
and the maximal size is $3.906$ at $\overline{\rho }=0$. Using this result,
the minimal work corresponding the thermal nucleation is 
\begin{equation}
W_{\min }=41.3376K_1\epsilon r_0^2h.  \eqnum{35}
\end{equation}
Comparing this with Eq. (29), we obtain the approximate formula for the
temperature characterizing the crossover from thermal to quantum nucleation
as 
\begin{equation}
k_BT_c\approx 0.55\frac{K_1\epsilon ^{1/4}}s\frac{\left| \cot \theta
_H\right| ^{1/6}}{\sqrt{1+\left| \cot \theta _H\right| ^{2/3}}}\left( \frac{%
1+\left| \cot \theta _H\right| ^{2/3}}{1-\epsilon +2\overline{K}_2\left(
1+\left| \cot \theta _H\right| ^{2/3}\right) }+\frac{2\chi _{\bot }K_1}{m^2}%
\right) ^{-1/2}.  \eqnum{36}
\end{equation}
For the FM case, i.e., the case of large non-compensation, the crossover
temperature is 
\begin{equation}
k_BT_c^{FM}\approx 0.55\frac{K_1\epsilon ^{1/4}}s\frac{\left| \cot \theta
_H\right| ^{1/6}}{\sqrt{1-\epsilon +2\overline{K}_2\left( 1+\left| \cot
\theta _H\right| ^{2/3}\right) }}.  \eqnum{37}
\end{equation}
While for the AFM case, i.e., the case of small non-compensation, 
\begin{equation}
k_BT_c^{AFM}\approx 0.39\hbar \gamma \sqrt{\frac{K_1}{\chi _{\bot }}}%
\epsilon ^{1/4}\frac{\left| \cot \theta _H\right| ^{1/6}}{\sqrt{1+\left|
\cot \theta _H\right| ^{2/3}}}.  \eqnum{38}
\end{equation}

To observe the quantum nucleation one needs a large crossover temperature
and not too small a nucleation rate. Note that $\chi _{\bot }=m_1^2/J_{ex}$
and $m_1=\hbar \gamma S$ where $J_{ex}$ is the exchange interaction between
two sublattices and $S$ is the sublattice spin, Eq. (36) can be written as 
\begin{eqnarray*}
k_BT_c &\approx &0.55\frac{K_1\epsilon ^{1/4}\hbar \gamma }m\frac{\left|
\cot \theta _H\right| ^{1/6}}{\sqrt{1+\left| \cot \theta _H\right| ^{2/3}}}
\\
&&\times \left( \frac{1+\left| \cot \theta _H\right| ^{2/3}}{1-\epsilon +2%
\overline{K}_2\left( 1+\left| \cot \theta _H\right| ^{2/3}\right) }+2\left( 
\frac Ss\right) ^2\left( \frac{K_1}{J_{ex}}\right) \right) ^{-1/2}.
\end{eqnarray*}
In Fig. 3, we plot the $\theta _H$ dependence of the crossover temperature $%
T_c$ for typical values of parameters for nanometer-scale antiferromagnets: $%
K_1=10^7$ erg/cm$^3$, $J_{ex}=10^{10}$ erg/cm$^3$, $\overline{K}_2=1$, $m=10$
emu/cm$^3$, $S/s=100$, $\epsilon =0.01$ in a wide range of angles $\pi
/2<\theta _H<\pi $. Fig. 3 shows that the maximal value of $T_c$ is about
1.916K at $\theta _H=2.350$, which is one or two orders of magnitude higher
than that for ferromagnets with a similar size.\cite{5,7} Note that, even
for $\epsilon $ as small as $10^{-3}$, the angle corresponding to an
appreciable change of the orientation of the N\'{e}el vector by quantum
tunneling is $\delta _2=\sqrt{6\epsilon }$ rad$>4^{\circ }$. The maximal
value of $T_c$ as well as $\Gamma _Q$ is expected to be observed in
experiment.

Now we study the situation that the magnetic field is applied opposite to
the initial easy axis, i.e., $\theta _H=\pi $. In this case, $\theta
_0=\theta _c=0$, $\overline{H}_x=0$, and $\eta =0$ from Eqs. (10a) and
(10b). By using the dimensional variables ${\bf r}^{\prime }=\epsilon ^{1/2}%
{\bf r}/r_0$, $\tau ^{\prime }=\epsilon ^{1/2}\omega _0\tau $, $\overline{%
\delta }=\delta /\sqrt{\epsilon }$, where $r_0=\sqrt{\alpha /2K_1}$, and 
\[
\omega _0=\frac{2\gamma K_1}m\left( \frac 1{2\overline{K}_2}+\frac{2\chi
_{\bot }K_1}{m^2}\right) ^{-1/2},
\]
we obtain the Euclidean action as 
\begin{eqnarray}
{\cal S}_E\left[ \overline{\delta }\left( {\bf r}^{\prime },\tau ^{\prime
}\right) ,\phi \left( {\bf r}^{\prime },\tau ^{\prime }\right) \right]  &=&%
\frac{r_0^3}{\epsilon ^2\omega _0}\int d\tau ^{\prime }d^3{\bf r}^{\prime
}\left\{ \frac{\chi _{\bot }}{2\gamma ^2}\epsilon ^2\omega _0^2\left[ \left( 
\frac{\partial \overline{\delta }}{\partial \tau ^{\prime }}\right) ^2+%
\overline{\delta }^2\left( \frac{\partial \phi }{\partial \tau ^{\prime }}%
\right) ^2\right] \right.   \nonumber \\
&&-i\frac m\gamma \epsilon ^{3/2}\omega _0\phi \overline{\delta }\left( 
\frac{\partial \overline{\delta }}{\partial \tau ^{\prime }}\right)
+2K_1\left[ \overline{K}_2^{\prime }\epsilon \overline{\delta }^2\phi
^2\right.   \nonumber \\
&&\left. \left. +\frac 12\epsilon ^2\left( \nabla ^{\prime }\overline{\delta 
}\right) ^2+\frac 12\epsilon ^2\overline{\delta }^2\left( \nabla ^{\prime
}\phi \right) ^2+\frac 12\epsilon ^2\left( \overline{\delta }^2-\frac{%
\overline{\delta }^4}4\right) \right] \right\} .  \eqnum{39}
\end{eqnarray}
The Gaussian integration over $\phi $ reduces Eq. (39) to the following
effective action 
\begin{equation}
{\cal S}_E^{eff}\left[ \overline{\delta }\left( {\bf r}^{\prime },\tau
^{\prime }\right) \right] =\frac{2K_1r_0^3}{\omega _0}\int d\tau ^{\prime
}d^3{\bf r}^{\prime }\left[ \frac 12\left( \frac{\partial \overline{\delta }%
}{\partial \tau ^{\prime }}\right) ^2+\frac 12\left( \nabla ^{\prime }%
\overline{\delta }\right) ^2+\frac 12\left( \overline{\delta }^2-\frac{%
\overline{\delta }^4}4\right) \right] .  \eqnum{40}
\end{equation}

In the case of quantum nucleation of the N\'{e}el vector in a small particle
of volume $V\ll r_0^3$, the N\'{e}el vector is uniform within the particle
and Eq. (40) reduces to 
\begin{equation}
{\cal S}_E^{eff}\left[ \overline{\delta }\left( \tau ^{\prime }\right)
\right] =\frac{2K_1}{\omega _0}\epsilon ^{3/2}V\int d\tau ^{\prime }\left[ 
\frac 12\left( \frac{\partial \overline{\delta }}{\partial \tau ^{\prime }}%
\right) ^2+\frac 12\left( \overline{\delta }^2-\frac{\overline{\delta }^4}4%
\right) \right] .  \eqnum{41}
\end{equation}
The classical trajectory satisfies the equation of motion 
\begin{equation}
\frac{d^2\overline{\delta }}{d\tau ^{\prime 2}}=\overline{\delta }-\frac 12%
\overline{\delta }^3,  \eqnum{42}
\end{equation}
which has the instanton solution 
\begin{equation}
\overline{\delta }\left( \tau ^{\prime }\right) =\frac 2{\cosh \tau ^{\prime
}},  \eqnum{43}
\end{equation}
corresponding to the variation of $\delta $ from $\delta =0$ at $\tau
=-\infty $, to $\delta =2\sqrt{\epsilon }$ at $\tau =0$, and then back to $%
\delta =0$ at $\tau =\infty $. Substituting this solution into Eq. (41) we
obtain that 
\begin{equation}
{\cal S}_{cl}=\frac 83\hbar s\epsilon ^{3/2}V\left( \frac 1{2\overline{K}_2}+%
\frac{2\chi _{\bot }K_1}{m^2}\right) ^{1/2}.  \eqnum{44}
\end{equation}
In the case of quantum nucleation in a thin film of thickness $h$ less than
the size $r_0/\sqrt{\epsilon }$ of the critical nucleus with its plane
perpendicular to the initial easy axis, the Euclidean action Eq. (39)
becomes 
\begin{equation}
{\cal S}_E^{eff}\left[ \overline{\delta }\left( {\bf r}^{\prime },\tau
^{\prime }\right) \right] =4\pi \hbar \epsilon ^{1/2}r_0^2h\left( \frac 1{2%
\overline{K}_2}+\frac{2\chi _{\bot }K_1}{m^2}\right) ^{1/2}\int duu^2\left[ 
\frac 12\left( \frac{d\overline{\delta }}{du}\right) ^2+\frac 12\left( 
\overline{\delta }^2-\frac{\overline{\delta }^4}4\right) \right] , 
\eqnum{45}
\end{equation}
where $u^2=\rho ^{\prime 2}+\tau ^{\prime 2}$, and $\rho ^{\prime
2}=x^{\prime 2}+y^{\prime 2}$. The corresponding classical equation of
motion satisfies 
\begin{equation}
\frac{d^2\overline{\delta }}{du^2}+\frac 2u\frac{d\overline{\delta }}{du}=%
\overline{\delta }-\frac 12\overline{\delta }^3.  \eqnum{46}
\end{equation}
The instanton solution of Eq. (46) can be found numerically and is
illustrated in Fig. 4. Numerical integration in Eq. (45), using this
solution, gives the WKB\ exponent for the subbarrier bubble nucleation in an
antiferromagnetic film as 
\begin{equation}
\Gamma _Q\varpropto \exp \left( -{\cal S}_E/\hbar \right) =\exp \left\{
-37.797s\epsilon ^{1/2}r_0^2h\left( \frac 1{2\overline{K}_2}+\frac{2\chi
_{\bot }K_1}{m^2}\right) ^{1/2}\right\} .  \eqnum{47}
\end{equation}
For a cylindrical bubble, the effective potential of the thermal nucleation
is found to be 
\begin{equation}
U_{eff}=4\pi K_1\epsilon r_0^2h\int_0^\infty d\rho ^{\prime }\rho ^{\prime
}\left[ \frac 12\left( \frac{d\overline{\delta }}{d\rho ^{\prime }}\right)
^2+\frac 12\left( \overline{\delta }^2-\frac{\overline{\delta }^4}4\right)
\right] ,  \eqnum{48}
\end{equation}
wherein the shape of the critical nucleus corresponds to a saddle point of
this functional: 
\begin{equation}
\frac{d^2\overline{\delta }}{d\rho ^{\prime 2}}+\frac 1{\rho ^{\prime }}%
\frac{d\overline{\delta }}{d\rho ^{\prime }}=\overline{\delta }-\frac 12%
\overline{\delta }^3.  \eqnum{49}
\end{equation}
Eq. (49) can be solved by the numerical approach, and the solution is showed
in Fig. 5. Numerical integration of this solution in Eq. (48) gives $\Gamma
_T\varpropto \exp \left( -W_{\min }/k_BT\right) $ with $W_{\min
}=23.402K_1\epsilon r_0^2h$. Then the crossover temperature is found to be 
\begin{equation}
k_BT_c\approx 0.619\frac{K_1\epsilon ^{1/4}}s\left( \frac 1{2\overline{K}_2}+%
\frac{2\chi _{\bot }K_1}{m^2}\right) ^{-1/2}.  \eqnum{50}
\end{equation}

In conclusion, we have investigated the quantum nucleation of the N\'{e}el
vector in nanometer-scale antiferromagnets with biaxial symmetry in the
presence of an external magnetic field at arbitrary angle. By applying the
instanton method in the spin-coherent-state path-integral representation, we
obtain the analytical formulas for quantum reversal of the N\'{e}el vector
in small magnets and the numerical formulas for quantum nucleation in thin
antiferromagnetic film in a wide range of angles $\pi /2<\theta _H<\pi $,
and $\theta _H=\pi $ respectively. The temperature characterizing the
crossover from the quantum to thermal nucleation is clearly shown for each
case. Our results show that the rate of quantum nucleation and the crossover
temperature depend on the orientation of the external magnetic field
distinctly. When $\theta _H=\pi $, the magnetic field is applied
antiparallel to the anisotropy axis. It is found that even a very small
misalignment of the field with the above orientation can completely change
the results of tunneling rates. Another interesting conclusion concerns the
field strength dependence of the WKB\ exponent or the classical action in
the rate of quantum nucleation. We have found that in a wide range of
angles, the $\epsilon \left( =1-\overline{H}/\overline{H}_c\right) $
dependence of the WKB exponent or the classical action $S_{cl}$ is given by $%
\epsilon ^{5/4}$, not $\epsilon ^{3/2}$ for $\theta _H=\pi $. Therefore,
both the orientation and the strength of the external magnetic field are the
controllable parameters for the experimental test of quantum nucleation of
the N\'{e}el vector in nanometer-scale antiferromagnets. If the experiment
is to be performed, there are three control parameters for comparison with
theory: the angle of the external magnetic field $\theta _H$, the strength
of the field in terms of $\epsilon $, and the temperature $T$.

Recently, Wernsdorfer and co-workers have performed the switching field
measurements on individual ferrimagnetic and insulating BaFeCoTiO
nanoparticles containing about $10^5$-$10^6$ spins at very low temperatures
(0.1-6K).\cite{22} They found that above 0.4K, the magnetization reversal of
these particles is unambiguously described by the N\'{e}el-Brown theory of
thermal activated rotation of the particle's moment over a well defined
anisotropy energy barrier. Below 0.4K, strong deviations from this model are
evidenced which are quantitatively in agreement with the predictions of the
MQT theory without dissipation.\cite{19,23} It is noted that the observation
of quantum nucleation of ferromagnetic or antiferromagnetic bubbles would be
extremely interesting as the next example, after single-domain
nanoparticles, of macroscopic quantum tunneling. The experimental procedures
on single-domain ferromagnetic\ nanoparticles of Barium ferrite with
uniaxial symmetry\cite{22} may be applied to the antiferromagnetic systems.
Note that the inverse of the WKB exponent $B^{-1}$ is the magnetic viscosity 
$S$ at the quantum-tunneling-dominated regime $T\ll T_c$ studied by magnetic
relaxation measurements.\cite{2} Therefore, the quantum nucleation of the
antiferromagnetic bubbles should be checked at any $\theta _H$ by magnetic
relaxation measurements. Over the past years a lot of experimental and
theoretical works were performed on the spin tunneling in molecular Mn$_{12}$%
-Ac\cite{24} and Fe$_8$\cite{25} clusters having a collective spin state $%
S=10$ (in this paper $S=10^6$). Recent progresses in the field of molecular
magnets include the measurements of the spin-lattice relaxiation rate,\cite
{30} the specific heat and the ac-susceptibility on Mn$_{12}$-Ac,\cite{31}
the studies of energy sublevels of the ground state by inelastic neutron
scattering (INS)\cite{32} and the zero-field magnetic relaxation of Mn$_{12}$%
-Ac,\cite{33} and the model calculation of the magnetization relaxation
based on phonon-assisted tunneling on Mn$_{12}$-Ac;\cite{34} the INS
experiment on Fe$_8$,\cite{35} the measurement of effects of nuclear spins
on the magnetic relaxation of Fe$_8$,\cite{36} and the nonadiabatic
Landau-Zener tunneling in Fe$_8$.\cite{37,38} Further experiments should
focus on the level quantization of collective spin states of $S=10^2$-$10^4$.

The ferromagnet or antiferromagnet is typically an insulating particle with
as many as $10^3\sim 10^6$ magnetic moments. For the dynamical process, it
is important to include the effect of the environment on quantum tunneling
caused by phonons, \cite{26,27} nucleation spins,\cite{28} and Stoner
excitations and eddy currents in metallic magnets.\cite{29} However, many
studies showed that even though these couplings might be crucial in
macroscopic quantum coherence, they are not strong enough to make the
quantum tunneling unobservable.\cite{2,26,27,28,29} The theoretical
calculations performed in this paper can be extended to the
antiferromagnetic systems with more general symmetries, such as trigonal,
tetragonal and hexagonal symmetries. We hope that the theoretical results
presented in this paper may stimulate more experiments whose aim is
observing quantum nucleation in nanometer-scale magnets.

\section*{Acknowledgments}

R. L. and Y. Z. would like to acknowledge Dr. Su-Peng Kou, Dr. Hui Hu, Dr.
Jian-She Liu, Professor Zhan Xu, Professor Mo-Lin Ge, Professor Jiu-Qing
Liang and Professor Fu-Cho Pu for stimulating discussions. R. L. and J. L.
Z. would like to thank Professor W. Wernsdorfer and Professor R. Sessoli for
providing their paper (Ref. 16). R. L. would like to thank Professor G. -H.
Kim for providing his paper (Ref. 21).

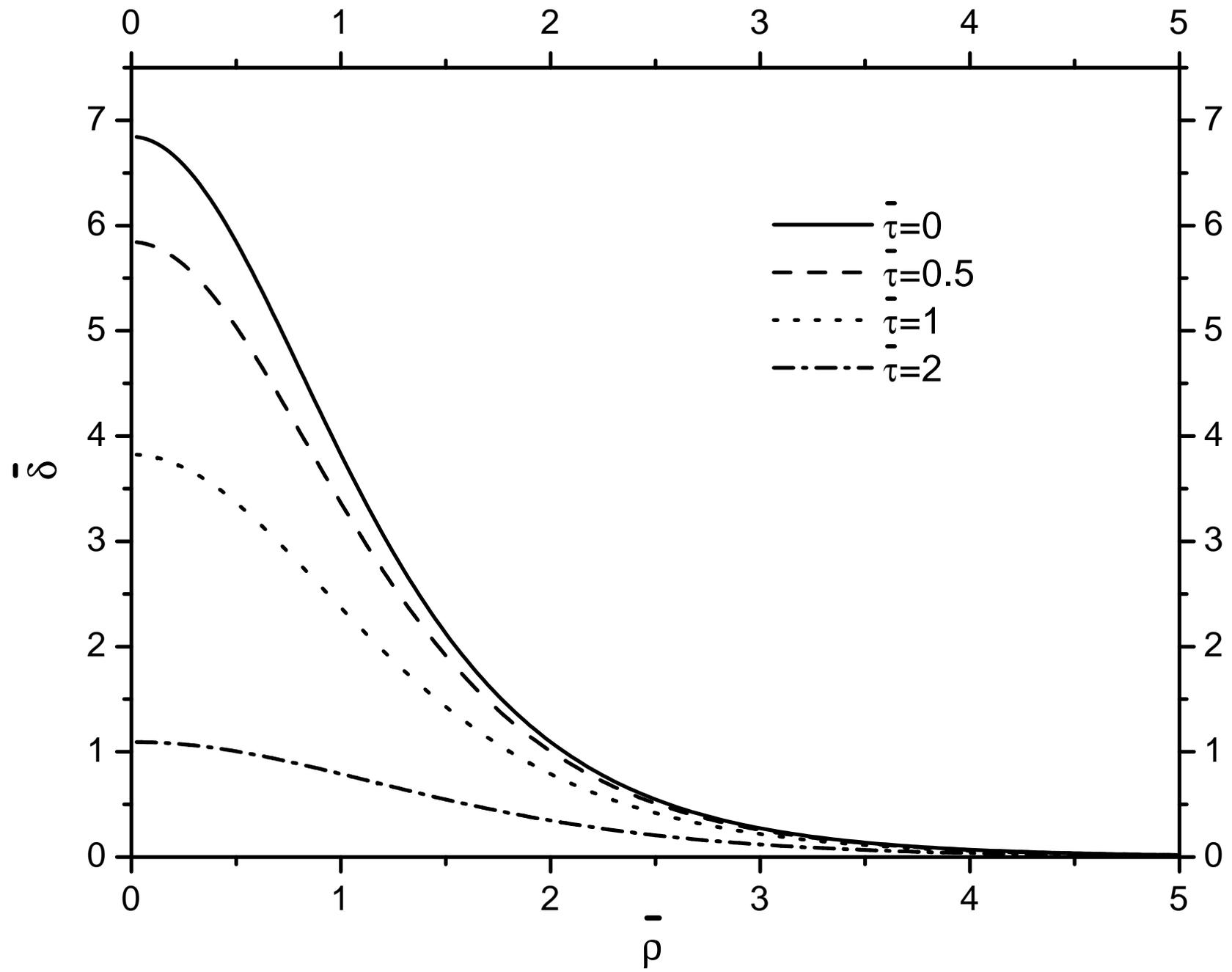
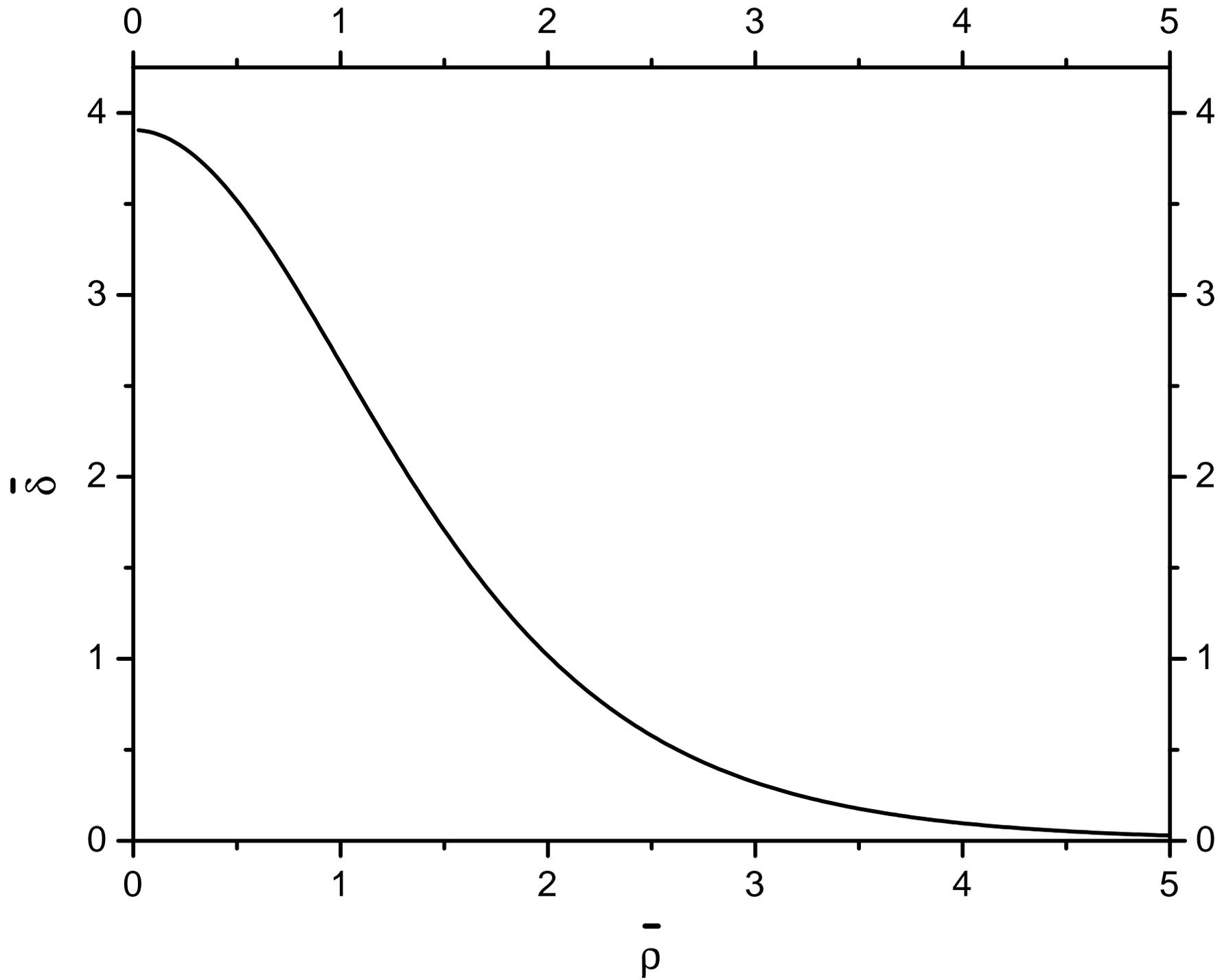
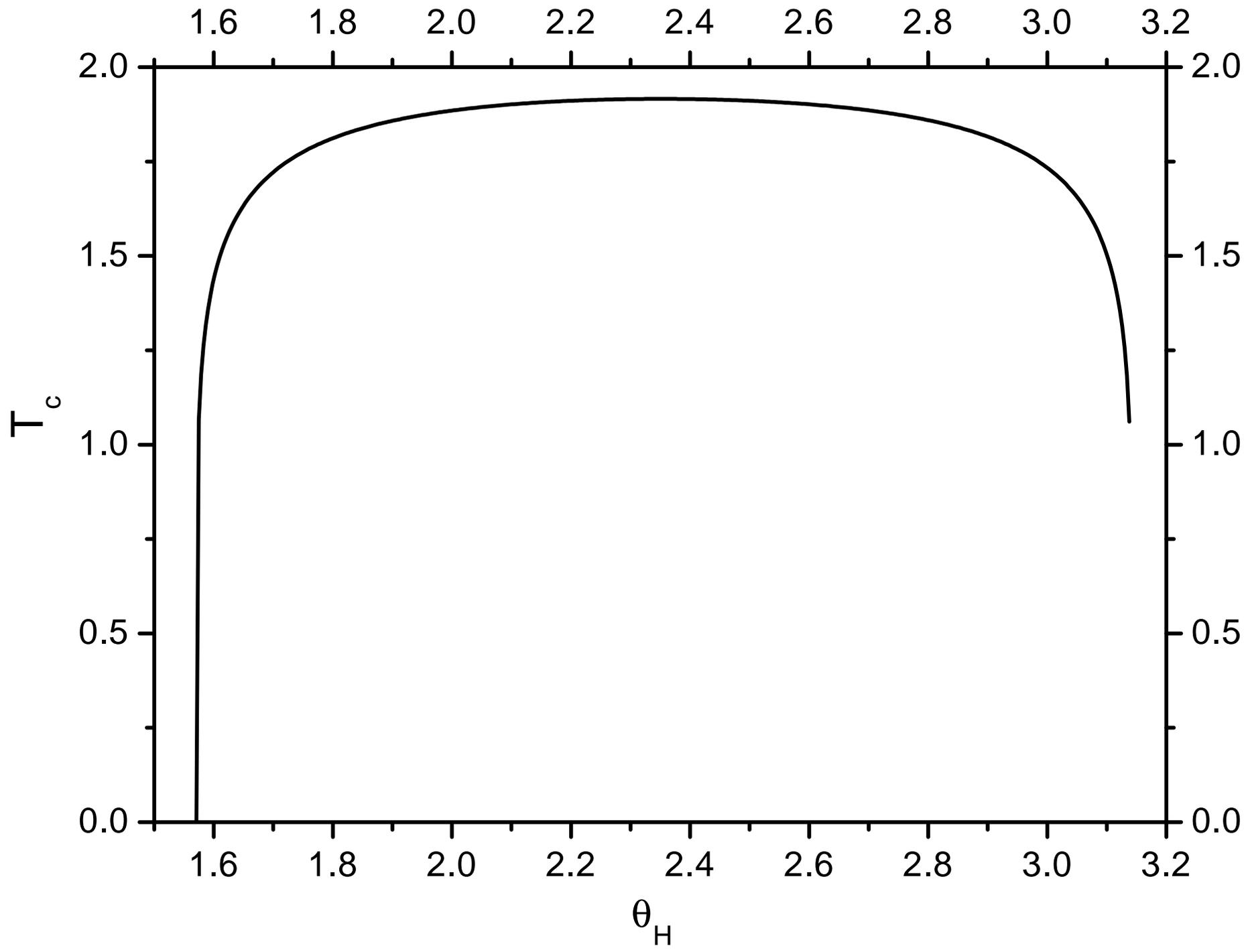
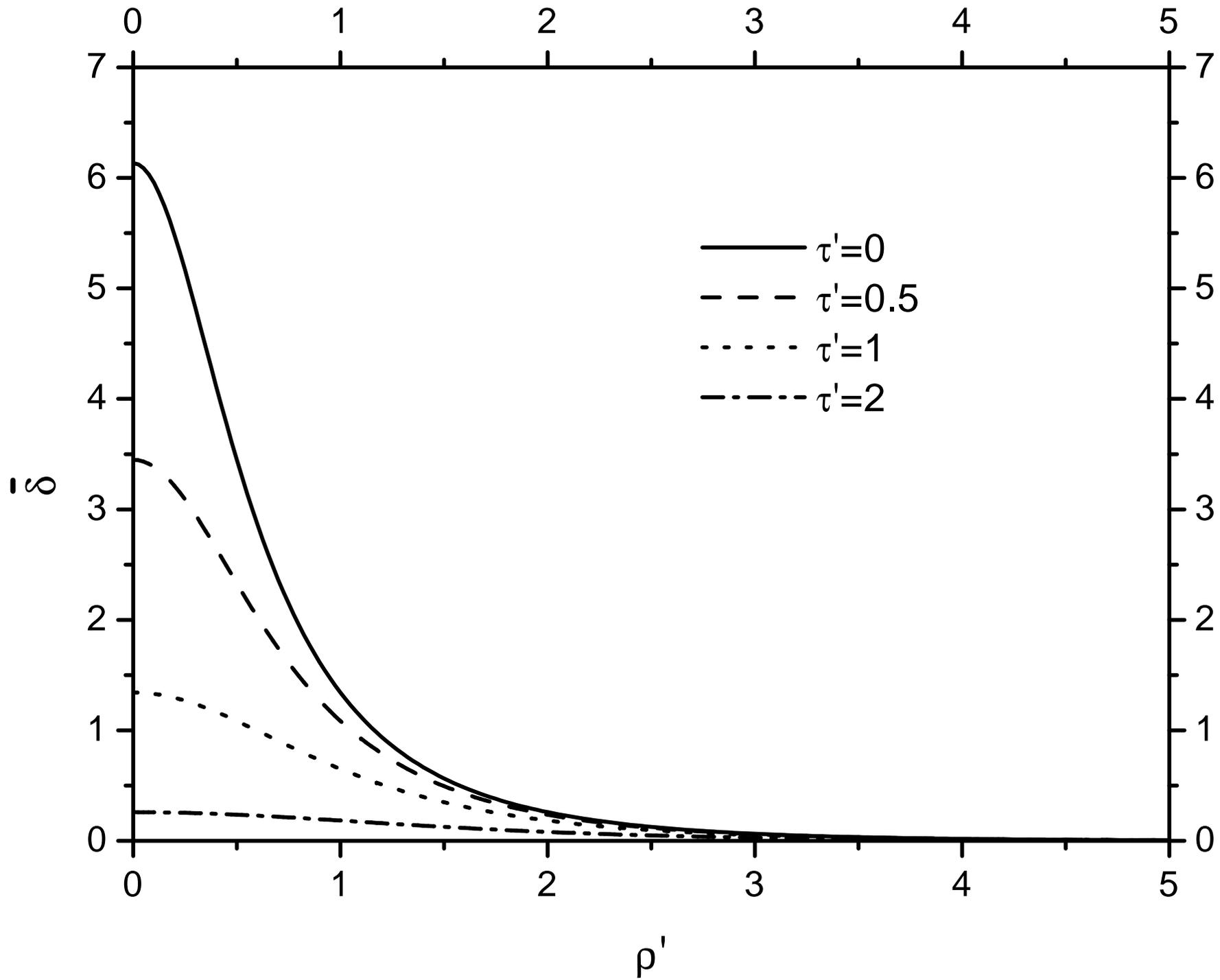
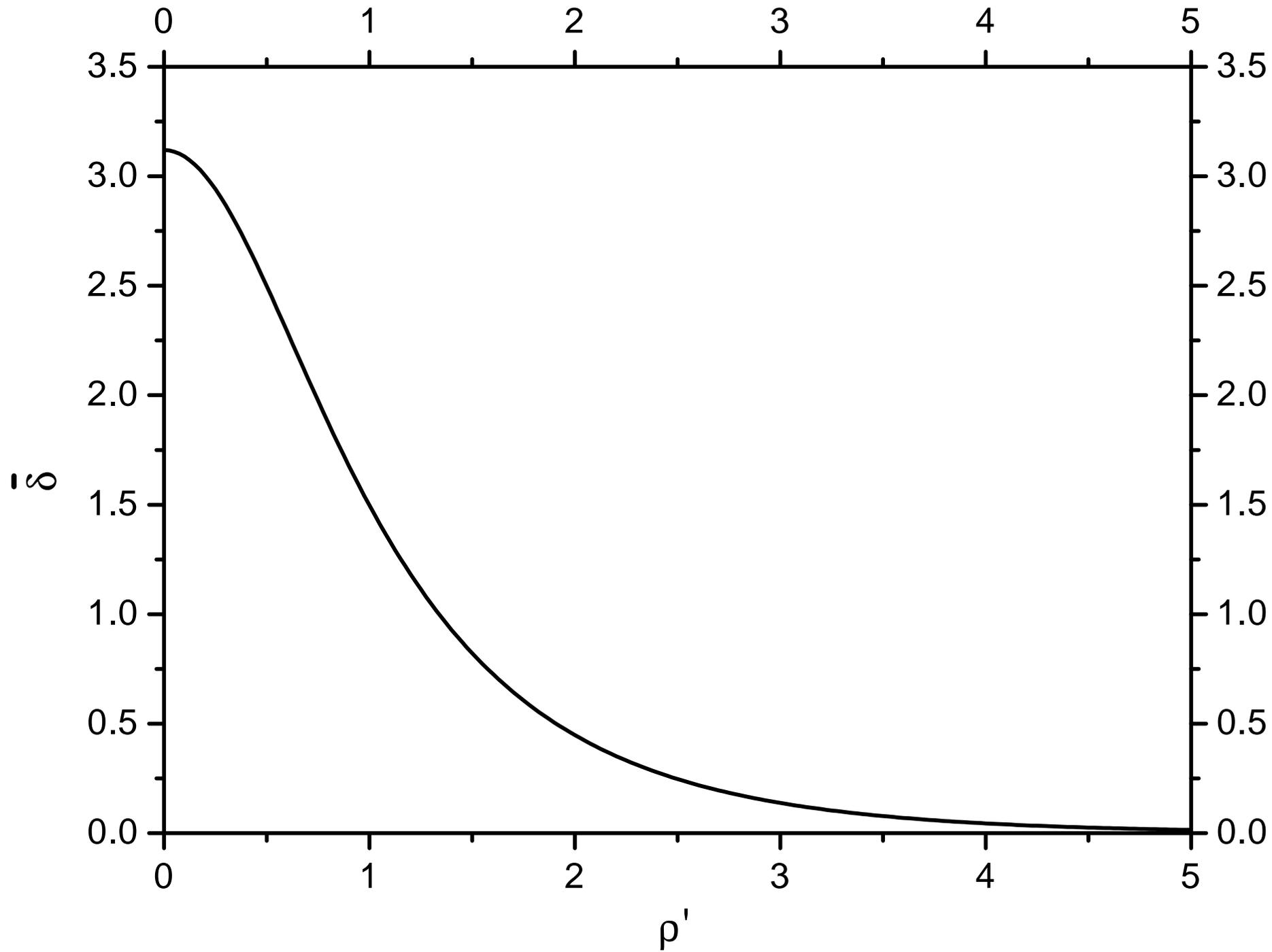
\end{document}